\documentclass[prl, twocolumn, showpacs, amsmath, amssymb , noshowkeys]{revtex4}

\usepackage{dcolumn}
\usepackage{bm}
\RequirePackage[usenames,dvipsnames]{color}

\usepackage{color}
\usepackage{epsfig} 
\psfull                                
\usepackage{mathbbol}
\usepackage{bbm}
\usepackage{dsfont}

\newcommand{\bra}[1]{\langle #1|}	
\newcommand{\ket}[1]{|#1\rangle}

\newcommand{\rme}{\text{e}}
\newcommand{\rmi}{\text{i}}
\newcommand{\sign}[1]{\text{sign}\left(#1 \right)}
\newcommand{\fh}{\mathcal{H}}
\newcommand{\brafl}[1]{\langle \langle #1|}	
\newcommand{\ketfl}[1]{|#1\rangle \rangle}

\newcommand{\eps}{\varepsilon}
\newcommand{\DDelta}[1]{\Delta_{#1}}

\newcommand{\ketT}[1]{|#1)}

\newcommand{\braketT}[2]{(#1|#2)}
\newcommand{\half}{\frac{1}{2}}
\newcommand{\omegaex}{\omega_\textrm{ex}}
\newcommand{\DXi}[2]{\Xi_{#2}^{#1}}
\newcommand{\EOU}[2]{E^{\uparrow}_{#1,#2}}
\newcommand{\EOD}[2]{E^{\downarrow}_{#1,#2}}

\newcommand{\ketTHD}[2]{\ketfl{\widetilde{\downarrow,#1,#2}}}

\newcommand{\Min}[1]{\text{Min} \left \{ #1 \right\} }

\usepackage{hyperref}

\begin{document}

\title{Qubit-oscillator system under ultrastrong coupling and extreme driving}
\date{\today}
\author{Johannes Hausinger} 
\email{johannes.hausinger@physik.uni-r.de}
\author{Milena Grifoni}
\affiliation{Institut f\"{u}r Theoretische Physik, Universit\"at
Regensburg, 93040 Regensburg, Germany}

\begin{abstract}
 We introduce an approach to studying a driven  qubit-oscillator system  in the ultrastrong coupling regime, where the ratio $g/\Omega$ between coupling strength  and oscillator frequency  approaches unity or goes beyond, and simultaneously for driving strengths much bigger than the qubit energy splitting (extreme driving). Both qubit-oscillator coupling and external driving lead  to a dressing of the qubit tunneling matrix element of different nature: the former can be used to suppress selectively certain oscillator modes in the spectrum, while the latter can bring the qubit's dynamics to a standstill at short times (coherent destruction of tunneling) even in the case of ultrastrong coupling. 
\end{abstract}

\maketitle

\begin{figure}
\centering
 \includegraphics[width=.47\textwidth]{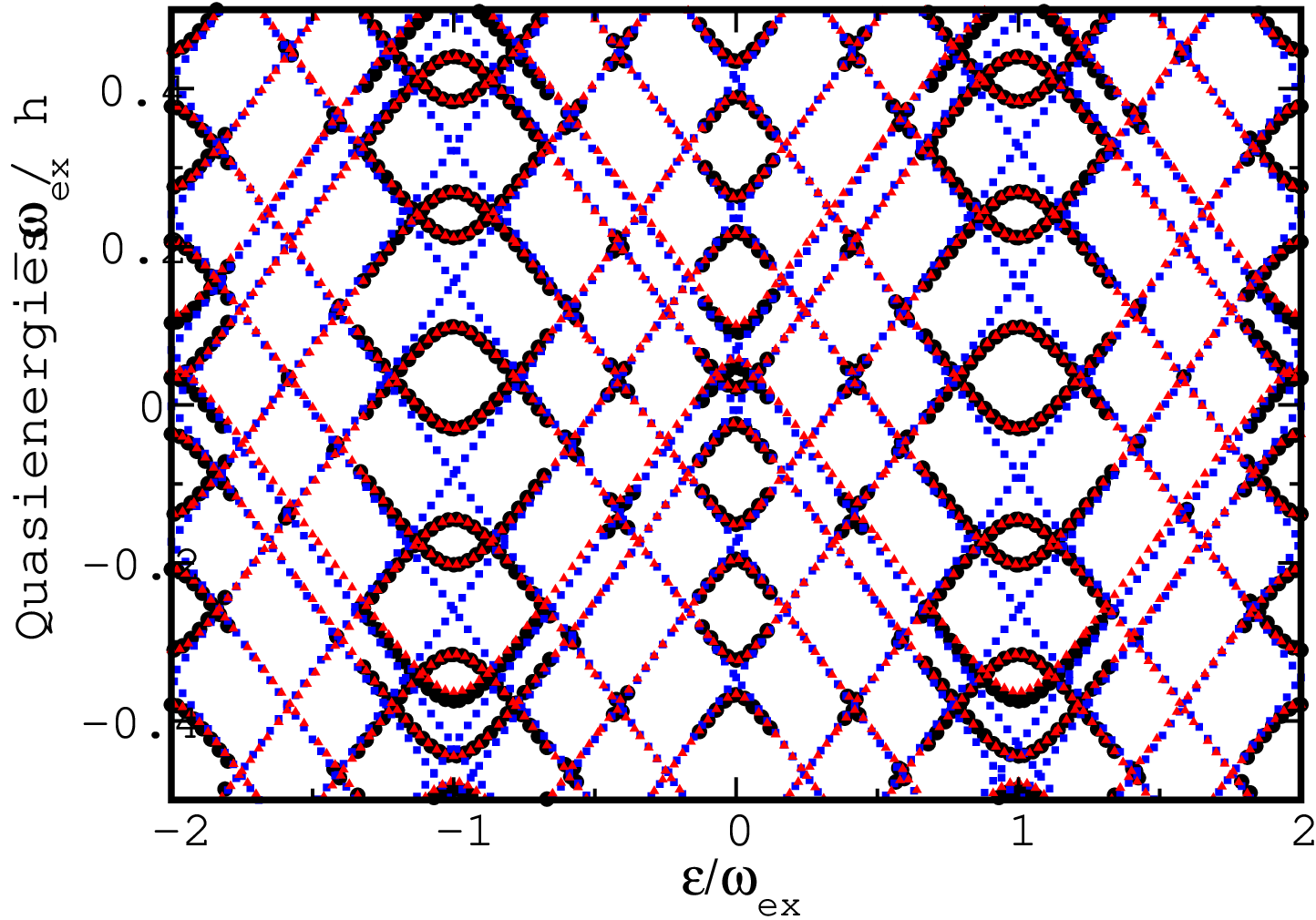}
\caption{(Color online) Quasienergy spectrum of the qubit-oscillator system  against the static bias $\eps$ for weak  coupling $g / \omegaex = 0.05$. Further parameters are  $\Delta/\omegaex=0.2$, $\Omega /\omegaex = \sqrt{2}$, $A /\omegaex = 2.0$. The first six oscillator states are included. Numerical calculations are shown by red (light gray) triangles, analytical results in the region of  avoided crossings by black dots. A good agreement between analytics and  numerics is found.  Blue (dark gray) squares represent the case $\Delta=0$. 
\label{Fig::QuasiEnEpsAnaDfinite}}
\end{figure}

\begin{figure}
\centering
 \includegraphics[width=8.3cm]{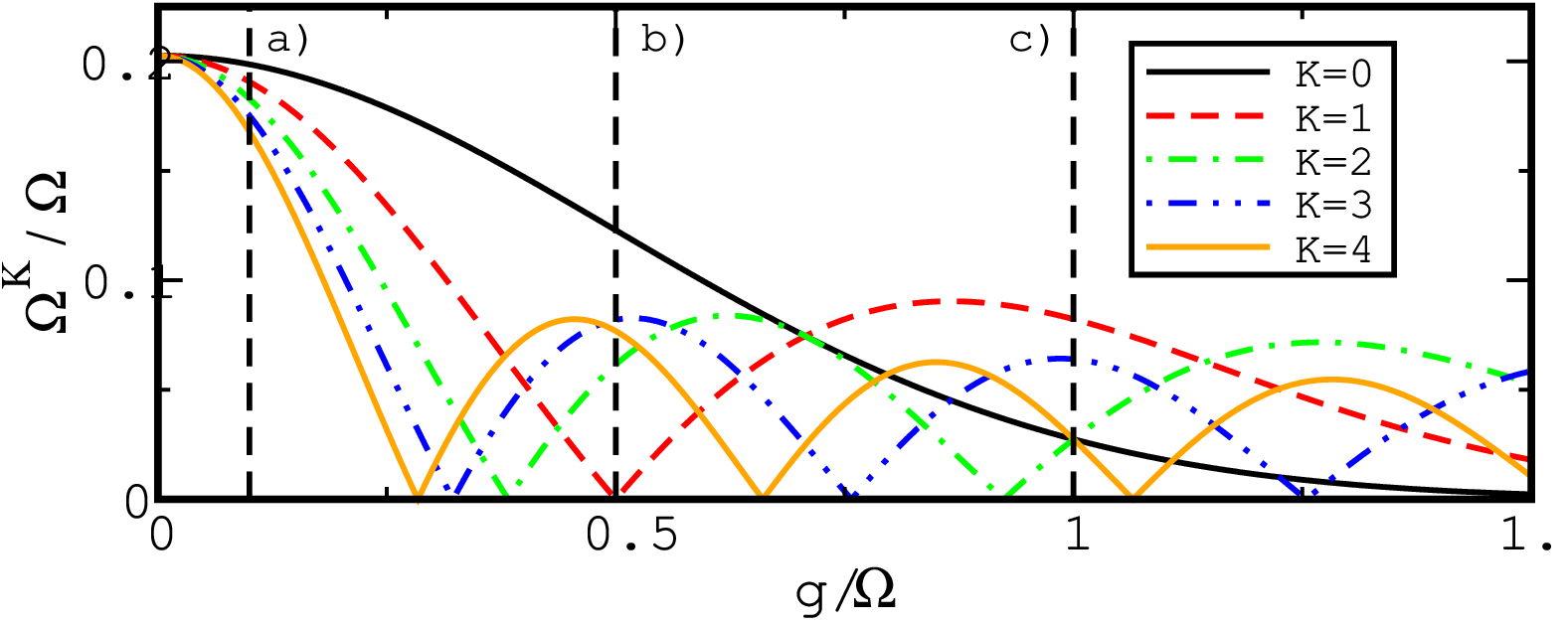}
\caption{(Color online) Size of the avoided crossing  $\Omega^{K}$ against the dimensionless coupling strength $g/\Omega$ for an unbiased qubit ($\eps=0$). Further, $\Delta/\Omega =0.4$, $\omegaex/\Omega =5.3$ and $A/\Omega =8.0$. $\Omega^K$ vanishes at the zeros of the Laguerre polynomial $L_K^0(\alpha)$. The dashed lines (a), (b), (c) represent $g/\Omega=0.1$, $0.5$, $1.0$, respectively, as considered in Fig. \ref{Fig::Dynam1}. \label{Fig::DressedOsc}}
\end{figure}

\begin{figure}
\centering
 \includegraphics[width=8.3cm]{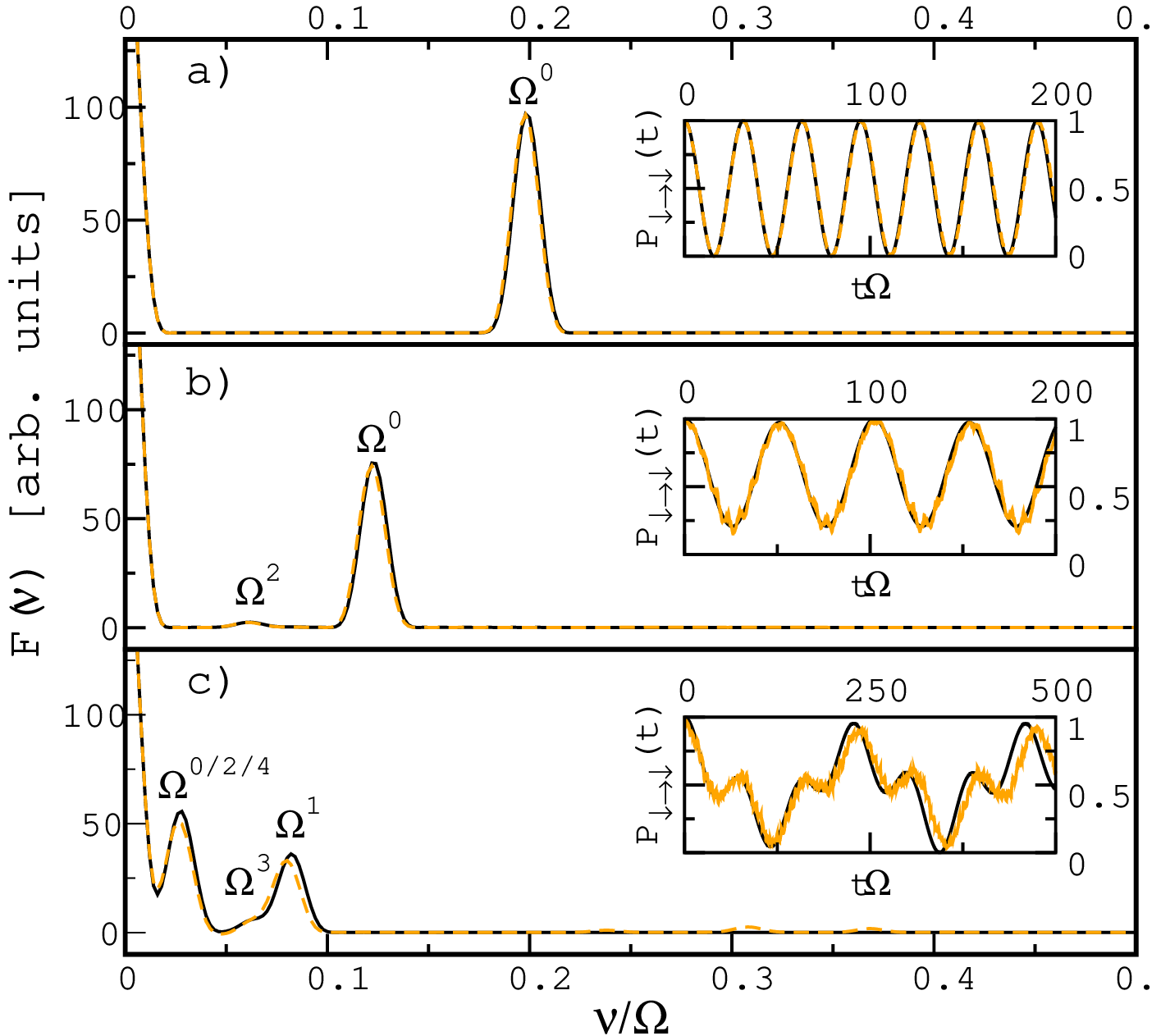}
\caption{(Color online) Dynamics of the qubit for $\eps=0$, $\Delta/\Omega =0.4$, $\omegaex/\Omega =5.3$, $A/\Omega =8.0$, and temperature $\hbar \Omega  (k_B T)^{-1} =10$.  The graphs show the Fourier transform $F(\nu)$ of the survival probability $P_{\downarrow \to \downarrow}(t)$ (see the insets). We study the different coupling strengths indicated in Fig. \ref{Fig::DressedOsc},  $g/\Omega=0.1$ (a), $0.5$ (b) and $1.0$ (c).
Analytical results are shown by black curves, numerics by  dashed orange curves. \label{Fig::Dynam1}}
\end{figure}

The model of a two-level system coupled to a harmonic oscillator has been a standard applied in many different fields of physics. For instance, in quantum optics it is used to describe the interaction between light and matter, leading to the field of cavity quantum electrodynamics (QED), where an atom interacts with the electromagnetic field of a resonator \cite{Raimond2001, Haroche2006}. In the regime of strong coupling, where  coherent exchange of excitations between atom and cavity is possible, those setups have become interesting for the field of quantum information with the atom being used as qubit and the cavity as information carrier. 
Additionally, the enormous progress in the field of circuit QED, where atom and cavity are replaced by superconducting circuits  \cite{Wallraff2004, Chiorescu2004, Johansson2006}, opens the door to the ultrastrong coupling regime \cite{Ciuti2005, Devoret2007,  Bourassa2009} with coupling strengths $g$ between qubit and oscillator which are of the order of the oscillator frequency $\Omega$ (typical values for cavity QED experiments are  $g/\Omega \approx 10^{-6}$). Experimental realizations beyond the strong coupling regime have recently been reported \cite{Niemczyk2010, Forn-Diaz2010}.
 The physics behind the qubit-oscillator system is usually analyzed within the Jaynes-Cummings model (JCM) \cite{Jaynes1963}, which relies on a rotating-wave approximation (RWA) with respect to  $g$  and provides
  deep insight into various effects of cavity and circuit QED. 
However, for the ultrastrong coupling regime the RWA  fails, and theories beyond the JCM are needed, see, e.g., \cite{Casanova2010}.
An external probing of the qubit, e.g., by microwave radiation, can be modelled by the driven JCM \cite{Alsing1992}, where a RWA is additionally invoked for the coupling between the qubit and the classical driving field, limiting the validity of the model to moderate driving amplitudes. It has been shown \cite{Berlin2004} that a strong external driving of the oscillator  makes an inclusion of  counter-rotating terms necessary even in the regime where the qubit's tunneling splitting equals the oscillator frequency ($\Delta = \Omega$) and for  couplings $g/\Omega \approx 0.1 $, parameters for which the JCM is commonly used in the undriven case. Similar effects are expected if instead of the cavity the atom is driven. Such extreme driving strengths have  already been experimentally realized \cite{Nakamura2001, Wilson2007, Wilson2010, Tuorila2010} leading to a dressed qubit state \cite{CohenTannoudji2004}.\newline
In this work we examine analytically the spectrum and dynamics of a system exposed to both ultrastrong coupling \textit{and} extreme driving. To go beyond common RWA schemes, the qubit is treated  within Floquet theory \cite{Grifoni1998}, while the coupling to the quantized field is included using a polaron transformation \cite{Plata1993}. 
We study various control possibilities of the qubit's dynamics, from tunneling suppression to selective frequency generation. \newline
\textit{The spectrum.} The Hamiltonian of the driven qubit-oscillator system  reads
\begin{equation}\label{HamDrivTLSHO}
 \hat H = - (\hbar/2) [\eps(t) \hat \sigma_z + \Delta \hat \sigma_x] + \hbar g \hat \sigma_z (\hat B^\dagger + \hat B) + \hbar \Omega \hat B^\dagger \hat B,
\end{equation}
where $\hat \sigma_i$ are the Pauli matrices, and $B^\dagger$, $B$ the creation and annihilation operators of the oscillator. A sinusoidal variation  of the static bias $\eps$ is given by
 $\eps(t) = \eps + A \cos \omegaex t$. 
In order to treat the time dependence, we analyze the Floquet Hamiltonian $\hat \fh = \hat H(t) - \rmi \hbar \partial_t$ and consider the extended Hilbert or Sambe space $\mathfrak{H} \otimes \mathfrak{T}$ \cite{Grifoni1998}, where $\mathfrak{H}$ is the Hilbert space of the undriven system and $\mathfrak{T}$ the space of the time-periodic functions. A basis of $\mathfrak{T}$ is provided by the vectors $\ketT{l}$, whereby $\braketT{t}{l}= \exp\{ - \rmi l \omegaex t\}$.
The eigenstates of the driven qubit for $\Delta = 0$ are \cite{Hausinger2010}:
\begin{equation} 
   \ketfl{u^0_{\uparrow/\downarrow,n}}  = \ket{\uparrow / \downarrow} \sum_l J_{\pm (n-l)} \left( A/2 \omegaex \right) \otimes \ketT{l},\label{BasisUHT2}
\end{equation}
with quasienergies $\hbar \varepsilon_{\uparrow/\downarrow, n}^0 = \mp \frac{\hbar}{2} \varepsilon - \hbar n \omegaex$ and the  Bessel function $J_n(x)$.
For no interaction between the qubit and the oscillator, $g=0$, this eigenbasis is easily extended to the full Hamiltonian $\hat \fh$ by
 $\ketfl{u^0_{\uparrow/\downarrow,n,K}} \equiv \ketfl{u^0_{\uparrow/\downarrow,n}}  \ket{K},$ 
with $\ket{K}$ being an eigenstate of the oscillator.   
For the coupled system ($g\neq0$),  the eigenstates can be found for $\Delta =0$ with the help of the polaron transformation 
$ \hat U= \exp\{g (\hat B-\hat B^\dagger) \hat \sigma_z/\Omega\}$ and are a combination of the Floquet states of the qubit and displaced oscillator states:
\begin{equation} \label{CoupledEigenstates}
  \ketfl{\widetilde{\uparrow/\downarrow, n, K}} = \exp \left\{ \pm [g (\hat B-\hat B^\dagger)]/\Omega  \right\} \ketfl{u^0_{\uparrow/\downarrow,n,K}}
\end{equation} 
with the quasienergies
\begin{equation}  \label{QuasienergiesD0}
  \hbar E^{\uparrow/\downarrow}_{n,K} = \mp \hbar\eps/2 -\hbar n\omegaex + \hbar K \Omega - \hbar g^2/\Omega.
\end{equation} 
For $\Delta = 0$, this result is analytically exact and treats the problem for \textit{arbitrary coupling strength $g$}.
Figure \ref{Fig::QuasiEnEpsAnaDfinite} shows the energy spectrum of Eq. (\ref{QuasienergiesD0}) for $\Delta =0$ (blue squares).
At
 $\eps = m \omegaex -L \Omega$
crossings occur;
the quasienergies $\EOD{n+m}{K+L}$ and $\EOU{n}{K}$ are degenerate, with the integer numbers $K+L =  0, 1, \ldots, \infty$ and $m, n= -\infty, \ldots, \infty$ \footnote{Additional crossings occur independent of $\eps$ if driving and oscillator frequency are commensurable, $\Omega/\omegaex = j/N$ with integers $j,N>0$, resulting in infinite many degenerate states. We avoid such a situation by choosing incommensurable frequencies or high values for $j$ and $N$, so that only high-photon processes are affected.}.
Note that for  $L\neq0$  there  are always $L$ nondegenerate levels. For $L>0$ those are the first $L$ spin-down states (positive slope), while for $L<0$ the first $L$ spin-up states (negative slope). At finite $\Delta$ avoided crossings occur in the energy spectrum  at the sites of the resonances (red triangles and black dots in Fig. \ref{Fig::QuasiEnEpsAnaDfinite}).\newline
 To explain the origin of these avoided crossings we express  $\hat \fh$ in the basis (\ref{CoupledEigenstates}) yielding the off-diagonal elements 
\begin{align} \label{DressedDeltaHO}
 & \tilde \Delta_{n,K}^{n^\prime, K^\prime}\equiv\brafl{\widetilde{\downarrow,n,K}}  \Delta \hat \sigma_x \ketfl{\widetilde{\uparrow, n^\prime, K^\prime}} \nonumber \\
       & = [\sign{K^\prime - K}]^{|K^\prime - K|} \DDelta{n^\prime-n} \DXi{|K^\prime -K|}{\Min{K,K^\prime}} (\alpha).
\end{align} 
The  dressing $\DDelta{m} =\Delta J_{m} \left(A/\omegaex \right)$ of the tunneling matrix element results from the external driving  \cite{Grifoni1998}, while  $\DXi{L}{K} (\alpha)= \alpha^{L/2} \sqrt{K!/(K+L)!} \,L^{(L)}_K (\alpha) \rme^{-\frac{\alpha}{2}}$ stems from the coupling to the oscillator \cite{Irish2007, Ashhab2010,Hausinger2010(2)} with  $L_K^{(L)} (x)$ being the $K$th generalized Laguerre polynomial and
  $\alpha \equiv \left( 2g/\Omega \right)^2$.
\newline
In order to calculate the energy spectrum for finite $\Delta$,
we make use of Van Vleck perturbation theory  in analogy to \cite{Hausinger2010} and \cite{Hausinger2010(2)}. To first order in $\Delta$, we only take into account  states degenerate for $\Delta =0$ in $\hat \fh$,  together with the matrix elements connecting them. Corrections from the remaining off-diagonal elements are calculated to second order. The resulting   effective Hamiltonian consists of 2$\times$2 blocks (without loss of generality we assume that $L\geq 0$):
\begin{equation} \label{HTLSHOeff}
  \hbar \left( \begin{array}{ c c }
  \EOU{n}{K} - \frac{1}{4} \eps^{(2)}_{\uparrow, n,K} &   \frac{(-1)^{L+1}}{2}  \DDelta{-m} \DXi{L}{K} (\alpha) \\
  \frac{(-1)^{L+1}}{2}  \DDelta{-m} \DXi{L}{K} (\alpha) &  \EOD{n+m}{K+L}+ \frac{1}{4} \eps^{(2)}_{\downarrow, n+m, K+L}
\end{array}
  \right),
\end{equation}
where we introduced the second-order corrections
\begin{equation}
 \eps^{(2)}_{\uparrow/\downarrow, n, K} \equiv \operatorname*{\sum_{\mathit{p}=-\infty}^\infty \sum_{\mathit{P}=-\mathit{K}}^\infty}_{\{p,P\} \neq \{-m, \pm L \} } \left(\tilde \Delta_{n- p, K+P}^{n,K}\right)^2/(\eps + p \omegaex \pm P \Omega).
\end{equation} 
The dressed tunneling matrix element in Eq. (\ref{HTLSHOeff}) determines to first order the width of the avoided crossings in Fig. \ref{Fig::QuasiEnEpsAnaDfinite}. Dominant crossings are found for $\eps = m \omegaex$, where the static bias is an integer multiple of the driving frequency, and thus $L=0$. That means that both states belong to the \textit{same} oscillator quantum number $K$, and the dressing contains a Laguerre polynomial of the kind $L^0_K(\alpha)$. 
For the 2$\times$2 block in Eq. (\ref{HTLSHOeff}) the  eigenvalues  to the eigenstates  $\ketfl{\Phi^{\mp,n,K}_{m,L}}$  are found easily:
\begin{align} \label{Quasienergies}
 \hbar &E^{\mp,n,K}_{m,L} =  (\hbar/2) \left[ -(2 n + m) \omegaex + (2K +L) \Omega \right. \nonumber  \\
               &  + (\eps^{(2)}_{\downarrow, n+m, K+L} - \eps^{(2)}_{\uparrow, n, K})/4-2g^2/\Omega  \mp \Omega^{n,K}_{m,L} \biggr],
\end{align} 
where the upper indices denote the state of the qubit, the Floquet mode and oscillator quantum number, while the lower indices stand for the resonance condition. The width of the avoided crossings is given by
\begin{align} \label{DressedOsc}
 \Omega^{n,K}_{m,L} = & \left\{\left[\eps - m \omegaex + L \Omega +(\eps^{(2)}_{\downarrow, n+m, K+L} +\eps^{(2)}_{\uparrow, n, K})/4\right]^2 \right. \nonumber \\
          &+ \left. [\DDelta{-m} \DXi{L}{K}(\alpha)]^2\right\}^\half,
\end{align}
the dressed oscillation frequency, 
which, together with Eq. (\ref{Quasienergies}), is one major result of this work.
For the $L$ nondegenerate spin-down states the quasienergies  and eigenstates are simply $\EOD{n}{K} + \frac{1}{4} \eps^{(2)}_{\downarrow, n, K}$ and $\ketTHD{n}{K}$.
\begin{figure}
\centering
 \includegraphics[width=8.3cm]{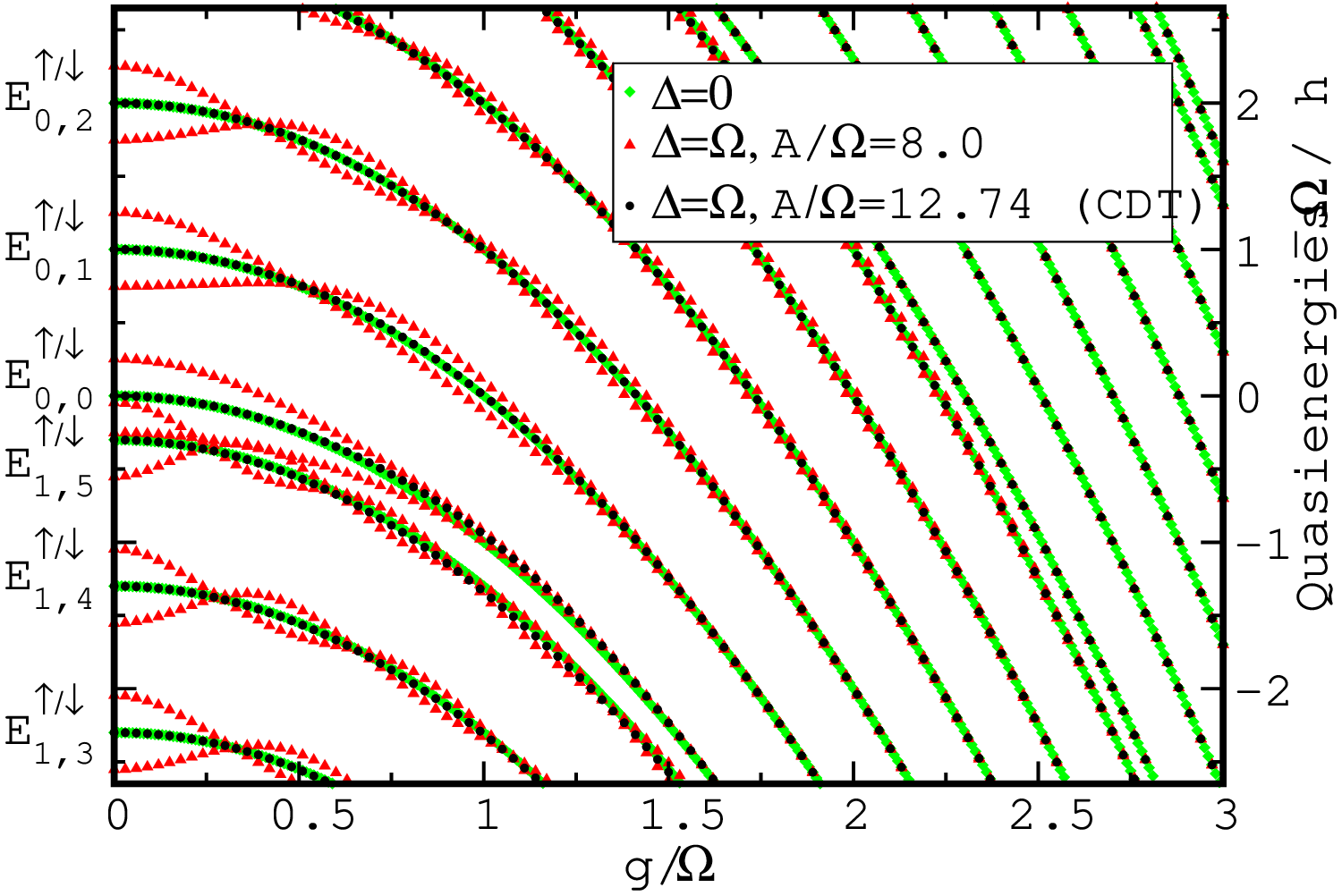}
\caption{(Color online) Quasienergy spectrum against the coupling strength $g$ in the unbiased case $\eps =0$. Further, we set $\omegaex/\Omega =5.3$ and  $\Delta/\Omega = 1.0$. The spectrum is examined for $A/\Omega = 8.0$  and $A/\Omega =12.74$ . For the former,  avoided crossings of amplitude $\Omega^K$ occur, which vanish at values of $g$ yielding zeros of the Laguerre polynomials. For the latter, all $\Omega^K$ vanish simultaneously for all values of $g$, since the CDT condition $J_0(A/\omegaex) = 0$ is fullfilled independently of the coupling strength $g$.  As a reference the  case $\Delta=0$ is shown. \label{Fig::SpectrumVSg}}
\end{figure}
\newline
Figure \ref{Fig::SpectrumVSg} shows the quasienergy spectrum against the  coupling strength $g$. For simplicity, we study the unbiased case $\eps =0$, which implies $m=L=0$ and hence gaps with   $\Omega^{n,K}_{0,0} = |\Delta_0L_K^0 (\alpha)\rme^{-\frac{\alpha}{2}}| \equiv \Omega^K$. 
Thus, for $g=0$ and $\Delta \neq 0$, the twofold degeneracy of the unperturbed case  is lifted by a gap of width $\Delta_0$. For $g\neq0$, the gap size is further determined by the Laguerre polynomial, so that additional degeneracies can occur at the zeros of $L_K^0 (\alpha)$.  When choosing the driving amplitude $A$ such that $\Delta_0=0$ the twofold degeneracy is kept for arbitrary $g$ and $K$. Because the dressing by the Bessel function does \textit{not} depend on $g$ or the oscillator level, we reach the remarkable conclusion that the coherent destruction of tunneling (CDT), predicted for a driven qubit \cite{Grossmann1992}, might occur also for a qubit-oscillator system in the ultrastrong coupling limit.
In Fig. \ref{Fig::DressedOsc}, the dressed oscillation frequencies are plotted against the dimensionless coupling $g/\Omega$. Next to an exponential decay, they exhibit zeros  that depend through the Laguerre polynomial characteristically on the oscillator quantum number $K$. Hence, because the qubit's dynamics involves several oscillator levels, we predict that suppression of tunneling cannot be reached by just tuning the coupling $g$.
\newline
\textit{The dynamics.} To prove the statements above, we calculate the survival probability of the qubit $P_{\downarrow \to \downarrow}(t) := \bra{\downarrow} \hat \rho_\text{red}(t) \ket{\downarrow}$, where  $\hat \rho_\text{red}$ is obtained by tracing out the oscillator degrees of freedom from the density operator of the qubit-oscillator system:
\begin{align} \label{DensityMatrix}
 \rho^{\alpha,K; \beta,K^\prime}_{m,L} (t) = & \bra{\Phi^{\alpha,K}_{m,L} (t) } \hat \rho(t) \ket{\Phi^{\beta,K^\prime}_{m,L} (t)} \nonumber \\
 = &\rho^{\alpha,K; \beta,K^\prime}_{m,L} (0)\, \rme^{-\rmi \omega_{m,L}^{\alpha, K;\beta,K^\prime}  t},
\end{align} 
with $\omega_{m,L}^{\alpha, K;\beta,K^\prime} := E^{\alpha,K}_{m,L} - E^{\beta,K^\prime}_{m,L}$ and $\{ \alpha, \beta \} \, \epsilon \, \{-, +,\downarrow \}$. The index $n$ has been dropped, because it  just leads to an overall phase  and thus has no influence on the dynamics.
The time-dependent Floquet modes $\ket{\Phi^{\alpha,K}_{m,L} (t)}$ can be determined from the eigenstates $\ketfl{\Phi^{\alpha,K}_{m,L}}$ of the effective Hamiltonian  (\ref{HTLSHOeff}) \cite{Hausinger2010}.
In Fig. \ref{Fig::Dynam1}, we plot the dynamics for zero bias and three different coupling strengths, indicated by the dashed lines in Fig. \ref{Fig::DressedOsc}. We assume factorized starting conditions for $t=0$, with the qubit in the state $\ket{\downarrow}$, and the oscillator in thermal equilibrium obeying a Boltzmann distribution. From Eq. (\ref{DensityMatrix}), one expects two main oscillatory contributions, namely, $\omega^{\mp K ; \pm K}_{m,L} = \pm \Omega^K_{m,L}$  and $\omega^{\alpha K ; \alpha K^\prime}_{m,L} = (K - K^\prime) \Omega$. Also sums of both can occur. 
For  weak coupling  $g/\Omega=0.1$ (a), the analytical calculation shows oscillations between the states $\ket{\downarrow}$ and $\ket{\uparrow}$ with the single frequency $\Omega^0$.  For stronger coupling $g/\Omega =0.5$ (b), a second small peak at $\Omega^2$ occurs in the Fourier spectrum, whose effect on the survival probability is almost not visible. The peak at $\Omega^{1}$ is absent, because the corresponding Laguerre polynomial vanishes at this value exactly, see Fig. \ref{Fig::DressedOsc}. Correspondingly, we observe a tunneling \textit{reduction} compared to case (a).
In Fig. \ref{Fig::Dynam1}(c) we are with $g/\Omega =1.0$ already deep in the ultrastrong coupling regime. The frequency $\Omega^1$ is now different from zero, and additionally $\Omega^3$ appears. The lowest peak belongs to the frequencies $\Omega^0$, $\Omega^2$, and $\Omega^4$, which are equal for $g/\Omega =1.0$, see Fig. \ref{Fig::DressedOsc}. A complete population inversion again takes place. Our results are confirmed by numerical calculations. For $g = 0.5, 1.0$, the latter yield additionally fast oscillations with $\Omega$ and $\omegaex$. Furthermore,  $\Omega^1$ is shifted in Fig. \ref{Fig::Dynam1}(c) slightly to the left, so that concerning the survival probability the analytical and numerical curves get out of phase for longer times. To include also the oscillations induced by the driving and the coupling to the quantized modes,  connections between the degenerate subspaces need to be included in the calculation of the  eigenstates of the full Hamiltonian \cite{Hausinger2010, Hausinger2010(2)}.
\begin{figure}
\centering
 \includegraphics[width=8.3cm]{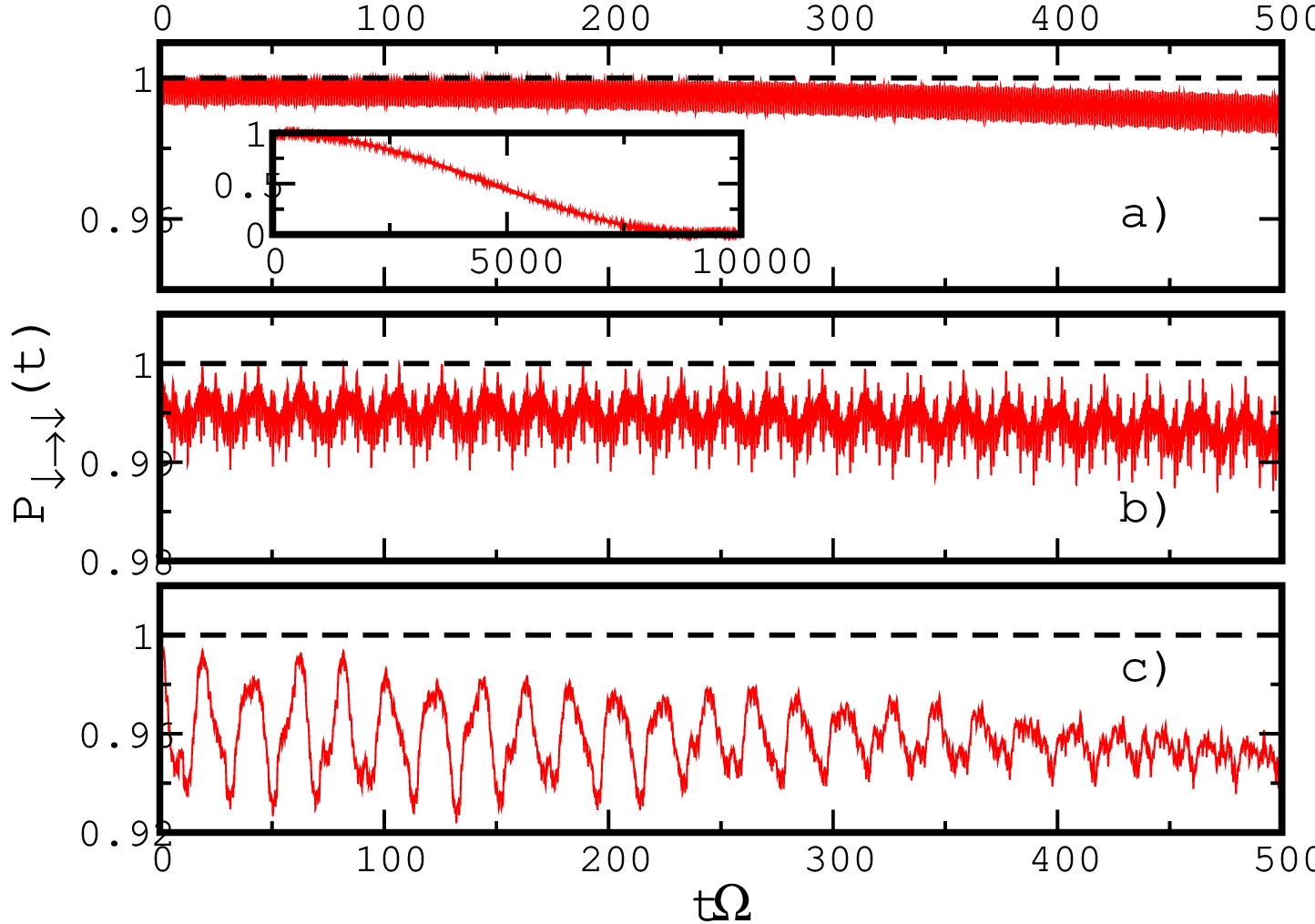}
\caption{(Color online) Coherent destruction of tunneling in a driven qubit-oscillator system. The same parameters as in Fig. \ref{Fig::Dynam1} are used except that  $A/\Omega =12.7$, which leads to $\Delta_0=0$. Three  coupling strengths are examined: $g/\Omega = 0.1$ (a), $0.5$ (b) and $1.0$ (c). The analytical calculations (black, dashed lines) predict complete localization for all three cases. Also the numerics (red curves) shows strong localization for short timescales with fast oscillations overlaid. For long times this localization vanishes (see inset in (a)). \label{Fig::DynamCDT}}
\end{figure}
\newline
While tuning the coupling $g$ to a zero of a Laguerre polynomial corresponding to a dominant oscillator mode yields a reduction of tunneling, tuning the driving amplitude $A$ to a zero of a Bessel function can yield almost complete localization at short times.
As  already noticed in Fig. \ref{Fig::SpectrumVSg}, this phenomenon is independent of the coupling strength $g$. We choose in Fig. \ref{Fig::DynamCDT} the driving amplitude $A$, so that $\Delta_0=0$. This is the same condition as found for CDT in a driven qubit \cite{Grossmann1992}. Analogously, our analytical solution now predicts localization for arbitrary coupling strength $g$. All dressed oscillation frequencies $\Omega^{K}$ vanish. However, third-order corrections in $\Delta$ will give small contributions to $\Delta_0$ \cite{Barata2000}. Hence, a numerical exact solution  yields oscillations of  $P_{\downarrow \to \downarrow}(t)$ with a long period.  On a short timescale and for $\omegaex \gg \Delta$ also the numerical solution appears to be strongly localized, while for long times, the inset in Fig. \ref{Fig::DynamCDT} (a) shows complete population inversion for the numerics.  \newline
In conclusions, we developed a powerful formalism to investigate analytically a qubit-oscillator system in the ultrastrong coupling \textit{and} extreme driving regime, a situation which is in close  experimental reach and offers excellent control possibilities.
Our approach relies on perturbation theory with respect to a single parameter only, the qubit tunneling matrix element $\Delta$, and thus goes
beyond the  driven Jaynes-Cummings model, with no rotating-wave approximation being applied.
\newline
We acknowledge financial support under DFG Program SFB631. We thank Sigmund Kohler for helpful remarks.


\end{document}